\documentclass[10pt]{article}
\usepackage[margin=1.2in]{geometry}

\usepackage{graphicx}
\usepackage{fmtcount}

\usepackage{multirow}

\usepackage[hyphens]{url}
\usepackage{hyperref}

\usepackage{array}
\usepackage{amsmath}
\usepackage{amssymb}

\usepackage{mathtools}

\usepackage{listings}
\usepackage{color}
\usepackage{booktabs}

\begin{document}

\title{Practically efficient methods for performing bit-reversed
  permutation in {\tt C++11} on the {\tt x86-64} architecture}

\author{Christian Knauth\\
Freie Universit\"at Berlin\\
Institut f\"{u}r Informatik
\and
Boran Adas\\
Freie Universit\"at Berlin\\
Institut f\"{u}r Informatik
\and
Daniel Whitfield\\
Freie Universit\"at Berlin\\
Institut f\"{u}r Informatik
\and
Xuesong Wang\\
Freie Universit\"at Berlin\\
Institut f\"{u}r Informatik
\and
Lydia Ickler\\
Freie Universit\"at Berlin\\
Institut f\"{u}r Informatik
\and
Tim Conrad\\
Freie Universit\"at Berlin\\
Institut f\"{u}r Informatik
\and
Oliver Serang\footnote{To whom correspondence should be addressed}\\
University of Montana\\
Department of Computer Science
}

\date{{\small \today}}

\maketitle

\begin{abstract}
\noindent The bit-reversed permutation is a famous task in signal
processing and is key to efficient implementation of the fast Fourier
transform. This paper presents optimized {\tt C++11} implementations
of five extant methods for computing the bit-reversed permutation:
Stockham auto-sort, naive bitwise swapping, swapping via a table of
reversed bytes, local pairwise swapping of bits, and swapping via a
cache-localized matrix buffer. Three new strategies for performing the
bit-reversed permutation in {\tt C++11} are proposed: an inductive
method using the bitwise XOR operation, a template-recursive closed
form, and a cache-oblivious template-recursive approach, which reduces
the bit-reversed permutation to smaller bit-reversed permutations and
a square matrix transposition. These new methods are compared to the
extant approaches in terms of theoretical runtime, empirical compile
time, and empirical runtime. The template-recursive cache-oblivious
method is shown to be competitive with the fastest known method;
however, we demonstrate that the cache-oblivious method can more
readily benefit from parallelization on multiple cores and on the GPU.
\end{abstract}

\section*{Introduction}
The classic Cooley-Tukey fast Fourier transform (FFT) works by
recursively reducing an FFT to two FFTs of half the size (in the case
of decimation in time these smaller FFTs are evaluated on the even
indices and odd indices of the original
array)\cite{cooley:algorithm}. Cooley-Tukey is extremely important for
its many uses in scientific computing: for example, the FFT permits
numeric computation of the convolution of two arrays of length $n$ in
$O(n \log(n))$ steps rather than the $O(n^2)$ required by the naive
convolution algorithm\cite{proakis:introduction}. The simplicity,
efficiency, and broad utility of Cooley and Tukey's work has led it to
be considered one of the most influential computer science methods of
the \ordinalnum{20} century\cite{cipra:best}.

The simplest way to compute the Cooley-Tukey FFT is to compute it
out-of-place and using a buffer of size $n$. Then, the values from the
original array can be copied into different indices of the buffer so
that the even indices of the array are copied to indices $0, 1, \ldots
\frac{n}{2}-1$ of the buffer and so that the odd indices of the array
are copied to indices $\frac{n}{2}, \frac{n}{2}+1, \ldots n-1$ of the
buffer. This buffered approach is commonly referred to as ``Stockham
auto-sort''\cite{cochran:fast}.

However, computing an in-place FFT (\emph{i.e.} computing an FFT by
overwriting the array evaluated, without requiring a result buffer) is
more challenging. The even-odd permutation performed by the Stockham
FFT is equivalent to an $\frac{n}{2} \times 2$ matrix
transposition. For example, if $n=8$, then the indices $[0, 1, 2, 3,
  4, 5, 6, 7]$ can be thought of as a $4 \times 2$ matrix, whose
transposition corresponds to the desired even-odd permutation:
\[ 
\left[
  \begin{matrix}
    0 & 1\\
    2 & 3\\
    4 & 5\\
    6 & 7
  \end{matrix}
\right]^T = 
\left[
  \begin{matrix}
    0 & 2 & 4 & 6\\
    1 & 3 & 5 & 7
  \end{matrix}
\right],
\]
where the even-index values are now in the first row and the odd-index
values are now in the second row. This transposition is the inverse of
the ``faro shuffle'' (also called ``perfect
shuffle'')\cite{sedgewick:algorithms}, wherein a deck of cards is cut
in half and then the two halves are interleaved. In-place FFT requires
these matrix transpositions to be performed in place (\emph{i.e.},
transposing the matrices while using substantially less than $n$
additional memory); however, in-place transposition of non-square
matrices is challenging because the destination index being written to
does not necessarily swap with the source value (as would be the case
with a square matrix). To transpose an $M \times N$ matrix using at
most $O(M + N)$ space, existing algorithms require runtime $\in
\Omega(M N \log(M N))$\cite{fich:permuting}. Furthermore, even if a
fast $O(n)$ algorithm were available to perform the even-odd
permutation without significant allocating of memory (perhaps
exploiting the fact that this is a special case of matrix
transposition, where the numbers of rows and columns are both powers
of two), performing this permutation separates the FFT of length $n$
from the recursive FFTs of length $\frac{n}{2}$; this can have a
negative influence and prevent the compiler from recognizing shared
code in the recursive calls, such as reused trigonometric
constants. FFT code that can be optimized by the compiler is highly
valued, and can produce a substantial speedup\cite{myrnyy:simple}.

One approach is to perform the even-odd permutations for all of the
recursive calls at the start of the FFT. Each even-odd permutation can
be thought of as making the least significant bit into the most
significant bit, and thus performing all of them sequentially is
equivalent to permuting by swapping the indices with their bitwise
reversed indices. The ``bit-reversed permutation'' of $[0, 1, 2, 3, 4,
  5, 6, 7]$ (or $[000, 001, 010, 011, 100, 101, 110, 111]$ in binary)
would be $[0, 4, 2, 6, 1, 5, 3, 7]$ (or $[000, 100, 010, 110, 001,
  101, 011, 111]$ in binary). Given a function {\tt rev} that reverses
integer indices bitwise, the bit-reversed permutation is fairly
straightforward to perform ({\bf
  Listing~\ref{alg:simple-permutation}}). Note that the bit-reversed
permutation can be performed in-place by simply swapping index and
reversed index pairs; this is less complicated than in-place
transposition of a non-square matrix, because we are guaranteed that
{\tt rev(rev(i)) = i} (in a manner similar to square matrix
transposition). After applying a bit-reversed permutation, the FFT can
be performed by simply invoking a slightly rewritten version of the
FFTs, which can now assume that at every level of recursion, the
even-odd permutation has already been performed.

\begin{footnotesize}
  \lstset{language=C++,
    basicstyle=\ttfamily,
    keywordstyle=\color{blue}\ttfamily,
    stringstyle=\color{red}\ttfamily,
    commentstyle=\color{magenta}\ttfamily,
    morecomment=[l][\color{magenta}]{\#},
    breaklines=true,
  }
\begin{lstlisting}[label={alg:simple-permutation},caption={{\bf Performing the bit-reversed permutation with the help of an external {\tt rev} function.} Let {\tt LOG\_N} be the problem size, which is a {\tt constexpr} (\emph{i.e.}, it is known a constant known at compile time). Note that the function {\tt rev} is templated to take the word size used for reversal.}]
void naive_bitwise_permutation(T*__ restrict const v) {
  constexpr unsigned long int N = 1ul << LOG_N;

  for (unsigned long index=1; index<(N-1); ++index) {
    unsigned long reversed = rev<LOG_N>(index);

    // Comparison ensures swap is performed only once per unique pair (otherwise, every index will be swapped and then swapped back to re-create the original array):
    if (index<reversed)
      std::swap(v[index], v[reversed]);
  }
}
\end{lstlisting}
\end{footnotesize}

Although some digital signal processors (DSPs) provide native bit
reversal operations, many modern desktop CPUs (including the {\tt x64}
CPUs at time of publication), do not have an opcode to perform the
{\tt rev} function; therefore, efficient functions to compute {\tt
  rev} are necessary. Bit reversal can of course be performed bitwise
in $\Theta(b)$, where $n = 2^b$ and where $b$ corresponds to {\tt
  LOG\_N} in the {\tt C++} code. Performing the full bit-reversed
permutation using this naive bit reversal method thus requires
$\Theta(n b) = \Theta(n \log(n))$ steps.

\paragraph{Bytewise bit reversal:}
Because the bit reversal function {\tt rev} is called at every index,
optimizing the {\tt rev} function can be quite beneficial for
performance. One way to perform bit reversal significantly faster than
the naive bitwise approach is to swap with byte blocks instead of
bits. This is efficiently accomplished by hard-coding an array of {\tt
  unsigned char} types, which contains 256 entries, one for each
possible byte. For any byte {\tt B}, the accessing the table at index
{\tt B} returns the bit-reversed value of that byte\cite{j:best,
  anderson:bit}. With this reversed byte table, the same approach as
the naive bitwise method can be used, requiring $\frac{b}{8}$ steps
instead of the $b$ steps required by the naive bitwise method. This
bytewise method is much faster in practice, even though its asymptotic
runtime, like the naive bitwise method, is still $\in \Theta(n
\log(n))$.

% segue into discussion of cache performance:
However, even if the bit reversal function {\tt rev} were natively
supported and ran in $O(1)$ CPU clock cycles, the non-sequential
memory accesses performed by the bit-reversal permutation do not cache
effectively using the standard hierarchical cache model (which loads
contiguous chunks into the cache every time a non-cached value is
fetched from the next layer of memory hierarchy), which is used by
{\tt x64} computers. This contrasts with the other FFT code, which
accesses memory in an almost perfectly sequential manner (and thus
caches very effectively). Thus as a result, the cost of computing a
large FFT can actually spend a significant percent of its time
performing the bit-reversed permutation.

% segue into pairwise bit reversal method
\paragraph{Reversing bitwise using pairs of bits:}
One method for achieving better cache performance proceeds bitwise but
working 2 bits at a time (exchanging the most significant and least
signifficant bits). Then, instead of computing the full bit-reversed
index and then swapping if {\tt index < rev(index)}, this paired
bitwise method performs multiple swaps, one after each pair of bits
are exchanged\cite{perez:place}. Thus a greater number of swaps on
the array are performed, but the swaps achieve better spatial
locality. Even though the blocks of memory accesseed are not
sequential, they are more contiguous than in the naive bitwise and
bytewise approaches. Despite the greater number of swap operations
performed on the array, the asymptotic runtime of this approach is no
different from the naive bit reversal approach, and is thus $\in
\Theta(n \log(n))$.

\paragraph{Bit reversal using a matrix buffer:}
Carter \& Gatlin proposed a cache-optimized method which uses a matrix
buffer. The principal idea is that the buffer is small enough to fit
completely within the {\tt L1} or {\tt L2} cache, and thus the buffer
can be accessed in either row-major or column-major order and still
achieve good cache performance. Given {\tt index = x y z} where the
bit strings {\tt x} and {\tt z} have the same size $\log(\sqrt{t})$,
their method uses a matrix buffer of size $\sqrt{t} \times \sqrt{t} =
t$\cite{carter:towards}.

Because {\tt rev(index) = rev(z) rev(y) rev(x)}, an out-of-place
method for performing the bit-reversed permutation would copy
$\forall$~{\tt x},~$\forall$~{\tt y},~$\forall$~{\tt z},~{\tt
  dest[rev(z)~rev(y)~rev(x)]~$\gets$~source[x~y~z]}. The method
(denoted COBRA in\cite{carter:towards}) proceeds in two steps: 

\noindent $\forall$~{\tt y},\\
\mbox{} \quad $\forall$~{\tt x},~$\forall$~{\tt z},~{\tt
  buff[rev(x)~rev(z)]~$\gets$~source[x~y~z]}\\
\mbox{} \quad $\forall$~{\tt x},~$\forall$~{\tt z},~{\tt
  dest[z~rev(y)~x]~$\gets$~buff[x~z]}.\\

\noindent This can be modified slightly:

\noindent $\forall$~{\tt y},\\
\mbox{} \quad $\forall$~{\tt x},~$\forall$~{\tt z},~{\tt
  buff[rev(x)~z]~$\gets$~source[x~y~z]}\\
\mbox{} \quad $\forall$~{\tt x},~$\forall$~{\tt z},~{\tt
  dest[rev(z)~rev(y)~x]~$\gets$~buff[x~z]}.\\

Although Carter \& Gatlin describe an out-of-place implementation
(meaning that it writes to {\tt dest} rather than modifying {\tt
  source}), it is possible to adapt the method to an in-place version
by swapping data with {\tt buff} rather than simply copying to or from
{\tt buff}. This requires a third looping step to propagate the
changes made to the buffer back into the array. Note that even this
``in-place'' variant of COBRA still requires a buffer, and so it is
not truly an in-place method.

Assuming a $\Theta(b)$ {\tt rev} method, the asymptotic runtime of
COBRA can be found as follows: The $\forall$~{\tt y} loop requires
$\frac{n}{t}$ steps, and the $\forall$~{\tt x} and $\forall$~{\tt z}
loops each require $\sqrt{t}$ steps. By caching bit-reversed values
(\emph{e.g.}, {\tt rev(y)}) as soon as they can be computed, the
runtime is
\[
\underbrace{\frac{n}{t} \cdot }_{\forall~{\tt y}} \left( \underbrace{ \log(\frac{n}{t}) }_{\mbox{\footnotesize Computing {\tt rev(y)}}} +~ \underbrace{2 \cdot}_{\mbox{\footnotesize Copy to/from {\tt buff}}} \underbrace{\sqrt{t}}_{\forall~{\tt x}} \left(
\underbrace{\log(\sqrt{t})}_{\mbox{\footnotesize Computing {\tt rev(x)} or {\tt rev(z)}}} + \underbrace{\sqrt{t}}_{\forall~{\tt z}} \right) \right).
\]
Asymptotically, this runtime is $\in \Theta\left( n + \frac{n}{t}
\log(\frac{n}{t}) \right)$. Note that using a large buffer
(\emph{i.e.}, $t \gg 1$), the runtime approaches $\Theta\left( n
\right)$; however, this is achieved by sacrificing the improved cache
performance that can be achieved when $t$ is small enough to fit into
the {\tt L1} or {\tt L2} cache. Conversely, when $n \gg t$, then the
runtime will be $\in \Theta(n \log(n))$. Unlike cache-oblivious
methods, which perform well for any hierarchical caches, for any
problem size $n$, the COBRA algorithm needs to optimize the parameter
$t$ for a particular architecture.

COBRA has previously been demonstrated to outperform a method from
Karp\cite{carter:towards}; before that, Karp's method had been shown
to outperform 30 other methods\cite{karp:bit}.\newline

% end of introduction:
In this paper we introduce three additional methods for performing the
bit-reversed permutation. We then compare the theoretical and
practical performance to existing methods by benchmarking fast
implementations of all methods in {\tt C++11}.

\section*{Methods}

\paragraph{Inductive XOR method for generating bit-reversed indices:}
We first propose a method that avoids calling the {\tt rev} function
altogether. This is acccomplished by inductively generating
bit-reversed indices without the use of bitwise reversing or bytewise
reversing with a reversed byte table.

Begin with the first index and its reversed value: {\tt index =
  rev(index) = 0}. The next index can be trivially computed as {\tt
  index+1}, but the next reversed index is found via {\tt
  rev(index+1)}, which will not necessarily equal {\tt
  rev(index)+1}. Rather than explicitly call a {\tt rev} function
(whether it be bitwise or bytewise), we avoid this by making use of
the bitwise XOR function. {\tt index} XOR {\tt index+1} reveals only
the bits that have changed by incrementing. Reversing both {\tt index}
and {\tt index+1} and computing {\tt rev(index)} XOR {\tt
  rev(index+1)} will be equivalent to computing {\tt rev(}{\tt index}
XOR {\tt index+1)}, because the bitwise XOR operation does not
exchange any information between bits. Furthermore, {\tt index} XOR
{\tt index+1} will reflect the fact that all differing bits reflect a
carrying operation in binary addition; therefore, {\tt index} XOR {\tt
  index+1} must be of the form {\tt 000\ldots 0111\ldots 1}. A bit
string of this form can be reversed efficiently by simply shifting
left by the number of leading zeros, thereby producing a bit string of
the form {\tt 1\ldots 1110\ldots 000}.

The number of leading zeros in a bit string of $b$ bits can be
computed trivially in $O(b)$ steps, but it can also be computed in
$O(\log(b))$ steps. An $O(\log(b))$ runtime is accomplished by
bitmasking with a bit string with the most significant $\frac{b}{2}$
equal to 1 and the least significant $\frac{b}{2}$ bits equal to 0,
thereby uncovering which half contains a 1 bit. The first half with a
1 can be iteratively subdivided to find the most significant 1 in
$O(\log(b))$ steps\cite{anderson:bit}.

The number of leading zeros can also be computed in $O(1)$ via the
integer $\log_2$: this can be performed by casting to a float and
bitmasking to retrieve the exponent; however, this approach only works
for $b \leq 24$\cite{anderson:bit}. For large problems with $b > 24$,
that shortcoming can be solved by casting to {\tt double}, although
casting to {\tt double} would be less efficient. Fortunately, even
though the {\tt x64} architecture does not have an opcode for
performing {\tt rev} in $O(1)$, it does feature an opcode for
computing the number of leading zeros. In {\tt g++} and {\tt clang}
this can be accessed via the {\tt \_\_builtin\_clzl} name (which
counts the leading zeros in an {\tt unsigned long int}).

\begin{figure}
\centering
\includegraphics[width=6in]{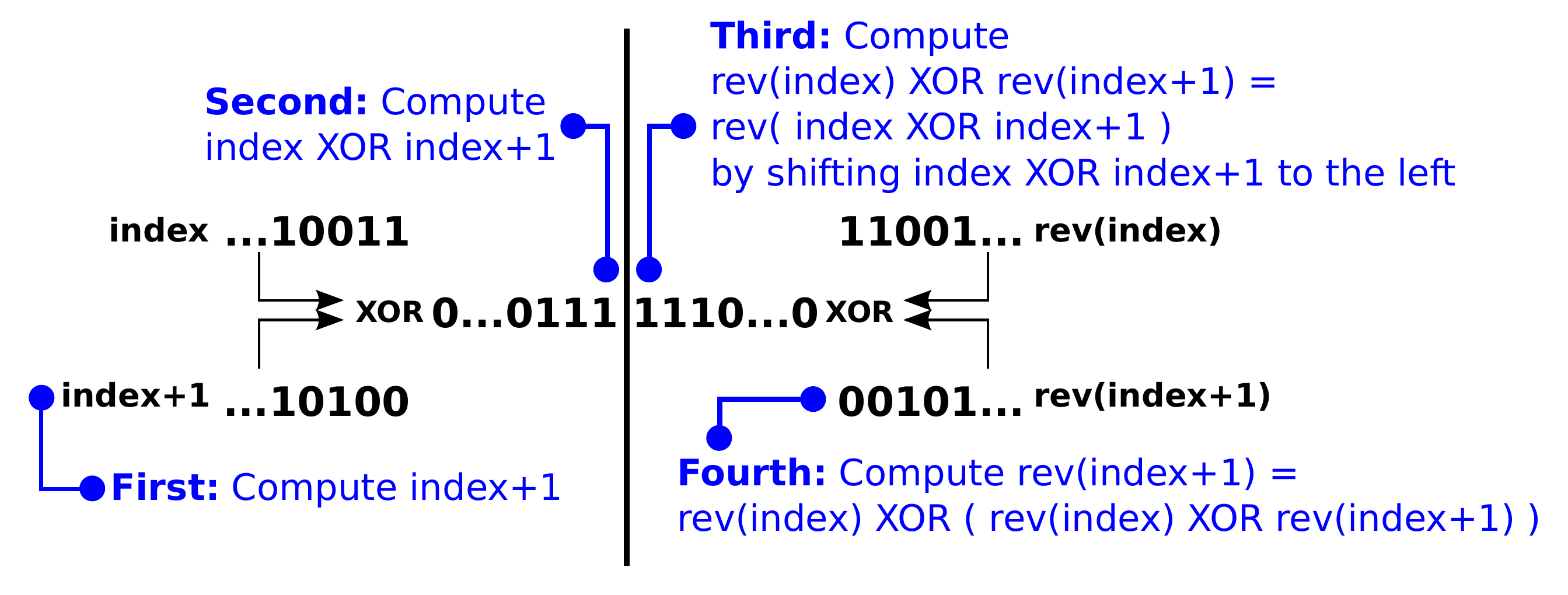}
\caption{{\bf Illustration of the inductive XOR method for bit
    reversal.} Beginning with {\tt index} and its bit-reverse value
  {\tt rev(index)}, the next values ({\tt index+1} and {\tt
    rev(index+1)}) are computed. First, {\tt index+1} is
  computed. Second, the bits that are different between {\tt index}
  and {\tt index+1} are computed via XOR; the resulting bit string
  will be of the form {\tt 000\ldots 0111\ldots 1}. Third, because the
  differing bits will be of the form {\tt 000\ldots 0111\ldots 1},
  they can be reversed by bit shifting left by the number of leading
  zeros. Fourth, {\tt rev(index+1)} can be computed by flipping only
  the differing bits (via XOR). 
  \label{figure:xor}}
\end{figure}

As a result, {\tt rev(index)} XOR {\tt rev(index+1)} can be reversed
in very few operations by computing the bits differing between {\tt
  rev(index)} and {\tt rev(index+1)}. The next reversed index, {\tt
  rev(index+1)}, is then computed by flipping only those bits
differing from {\tt rev(index)} by using XOR ({\bf
  Figure~\ref{figure:xor}}). The XOR method performs a constant number
of XOR operations as well as a single count leading zeros operation to
compute the next index and reversed index. Count leading zeros will
require $\in \Theta(\log(b)) = \Theta(\log(\log(n))$ operations in the
general case, and can be performed in $O(1)$ when $b \leq 24$ or when
hardware support a built-in operation. Thus the overall runtime of the
full bit-reversed permutation will be $\in \Theta(n \log(\log(n)))$ in
the general case and $\in \Theta(n)$ when count leading zeros can be
performed in $O(1)$.

The XOR method is very efficient at computing the next index and next
reversed index; however, even if that computation were instantaneous,
the algorithm achieves only moderate performance for two reasons: The
first reason is that the bit-reversed indices are not accessed in a
contiguous fashion (and thus are not cache-optimized). The second
reason is the {\tt if} statement, which is also found in {\bf
  Listing~\ref{alg:simple-permutation}}. This {\tt if} statement
prevents loop unrolling from being fully effective, because the
compiler will not yet know the effects of swapping in the previous
iteration (limiting the ability to swap in parallel, even if hardware
would support it). It should also be noted that the frequency with
which {\tt index < rev(index)} will change during the loop,
diminishing the benefits of branch prediction.

\paragraph{Unrolling (template-recursive closed form):}

Eliminating the {\tt if (index < rev(index))} comparison is difficult,
because it requires computing in advance the indices on which this
will be true and then only visiting those indices; correctly computing
the pattern of indices where this is true is strikingly similar to
performing bit reversal. However, for problems of a fixed size
(\emph{e.g.}, $b = 10$ bits or equivalently $n = 1024$), it is
possible to simply compute all indices that should be swapped over the
bit-reversed permutation. This offers two benefits: First, the
overhead of looping, performing the bit reversal on indices, and
checking whether {\tt index < rev(index)} are all eliminated. Second,
the compiler could (in theory) rearrange the swap operations to be
more sequentially contiguous, thereby improving cache performance.

\begin{figure}
\centering
\includegraphics[width=4.5in]{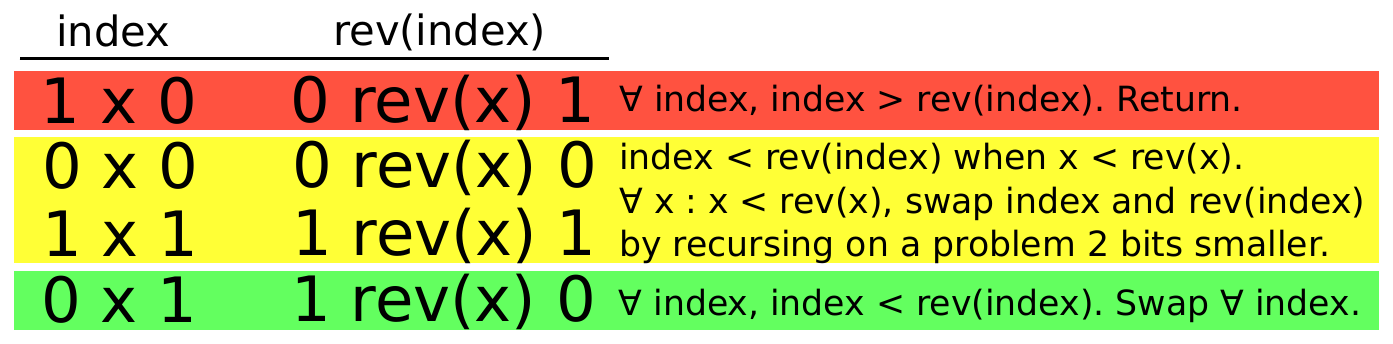}
\caption{{\bf Illustration of the unrolled method for bit reversal.}
  All possible bit strings where {\tt index < rev(index)} are found at
  compile time by beginning with the most significant bit remaining
  and least significant bit remaining and progressing inward
  recursively. Bit strings of the form {\tt 1~x~0} will never be less
  than their reverse (highlighted in red). Bit strings of the form
  {\tt 0~x~1} will always be less than their reverse (highlighted in
  green). Bit strings of the form {\tt 0~x~0} and {\tt 1~x~1} will
  only be less than their reverse when {\tt x < rev(x)} (highlighted
  in yellow).
  \label{figure:unrolled}}
\end{figure}

This could implemented via a hard-coded function {\tt
  unrolled\_permutation\_10}, but alternatively it could also be
implemented by generating that code at compile time via template
recursion. For any bit string of the form {\tt index = z x y} (where
{\tt z x y} is the concatenation of bit strings {\tt z}, {\tt x}, and
{\tt y}), the reverse will be {\tt rev(index) = rev(y) rev(x)
  rev(z)}. When the bit strings {\tt z} and {\tt y} consist of a
single bit, then their reversal can be ignored: {\tt rev(index) = y
  rev(x) z}. Thus, it is possible to start from both ends (the most
significant bit remaining and least significant bit remaining) and
proceed inward to recursively generate problems of the same form
(these recursive calls are implemented via template recursion to
unroll them at compile time). Some of the template recursions can be
aborted in a branch-and-bound style ({\bf
  Figure~\ref{figure:unrolled}}): Bit strings of the form {\tt index =
  1~x~0} will never be less than their reverse, regardless of the
value of the bit string {\tt x}; therefore, further recursion can be
aborted. Likewise, bit strings of the form {\tt index = 0~x~1} will
always be less than their reverse, and therefore swap operations
should be performed for every possible {\tt x}. Lastly, bit strings of
the form {\tt 0~x~0} and {\tt 1~x~1} will be less than their reverse
when {\tt x < rev(x)}. This produces a problem of the same form as
finding whether {\tt index < rev(index)}, but two bits smaller, and
thus it can be performed recursively.

As a result, the runtime of this method is defined by the recurrence
$r(b) = \underbrace{2^{b-2}}_{green} + \underbrace{2 \cdot
  r(b-2)}_{yellow} + \underbrace{0}_{red}$ (where the colors labeling
the three terms correspond to the coloring in {\bf
  Figure~\ref{figure:unrolled}}). Using $r(1) = 0$ (because there are
0 swaps necessary with a single bit) and $r(2) = 1$ (there is one swap
necessary using 2 bits), then this recurrence has closed form $r(b) =
2^{\frac{b}{2}} \cdot \left( {(-1)}^b \frac{\sqrt{2}-1}{4} - \frac{1 +
  \sqrt{2}}{4} \right) + 2^{b-1}$. Hence, regardless of whether $b$ is
even or odd, $r(b) = 2^{b-1} - c \cdot 2^{\frac{b}{2}}$
($c=\frac{-1}{2}$ when $b$ is even and $c=\frac{-1}{\sqrt{2}}$ when
$b$ is odd). Regardless, the asymptotic runtime is dominated by the
$2^{b-1}$ term; therefore, asymptotic runtime is a linear function of
$n$ (because $n = 2^b$).

For small problems, this template-recursive ``unrolled'' method is
very efficient. However, a disadvantage of the method is that on large
problems, a great deal of code will be generated, which can result in
large compilation times and also can mean that the order of the
swapping operations will not be optimized effectively by the
compiler. In fact, there may be no cache-efficient order to visit the
indices being swapped because bit reversed indices jump around. Thus,
unless the compiler possesses mathematical insight about the swaps
being performed (enabling the compiler to transform the code into one
of the other algorithms listed), the unrolled method's mediocre
efficiency on large problems will not justify its substantial
compilation times.

\paragraph{Cache-oblivious recursive bit-reversed permutation:}

When the number of bits $b$ is even, an index bit string can be
partitioned into two equally sized bit strings of $\frac{b}{2}$ bits
each: {\tt index = x~y}. The reversed index would be {\tt
  rev(index)~=~rev(y)~rev(x)}. This can be computed in three steps:
First, reversing the least significant bits (which form {\tt y}) to
produce {\tt x~rev(y)}. Second, swapping the most significant bits
forming {\tt x} with the least significant bits forming {\tt rev(y)},
thus producing {\tt rev(y)~x}. Third, repeating the first step and
reversing the least significant bits (which at this point contain {\tt
  x}) to produce {\tt rev(y)~rev(x)~=~rev(index)}.

\begin{figure}
\centering
\includegraphics[width=5in]{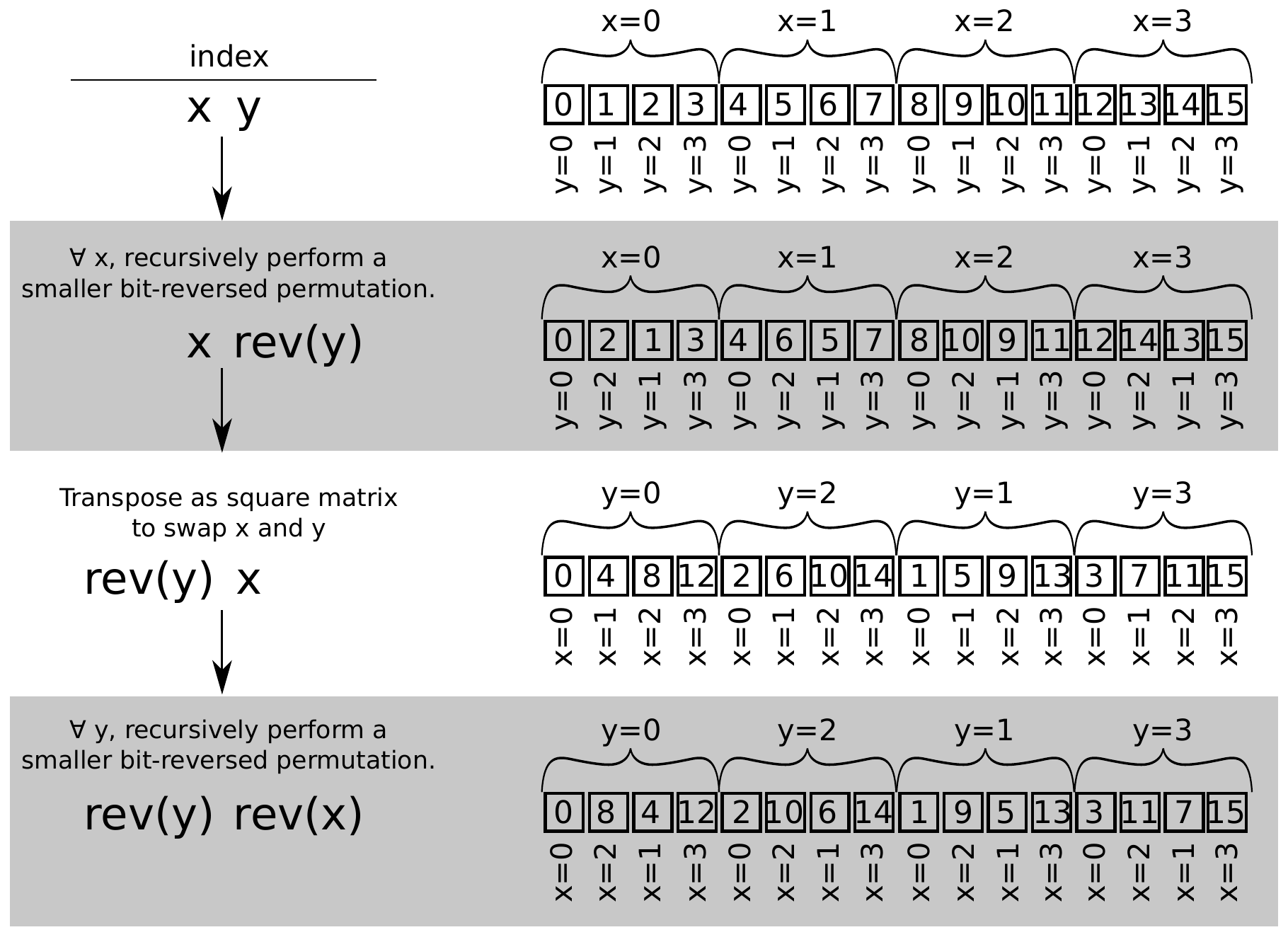}
\caption{{\bf Illustration of recursive bit-reversed permutation.} A
  bit-reversed permutation on $b$ bits is performed as several smaller
  bit-reversed permutations on $\frac{b}{2}$ bits, an in-place
  transposition of a square matrix and another batch of smaller
  bit-reversed permutations on $\frac{b}{2}$ bits. The spatial
  locality of each of these steps yields a cache-performant algorithm.
  \label{figure:recursive}}
\end{figure}

This recursive technique is not only useful for computing a reversed
index: it can also be used to perform the entire bit-reversed
permutation in a recursive manner with localized operations that do
not need to be tuned for the cache size. The first and third steps are
%fixme: really the most significant bits?
performed in a manner identical to one another: for all possible most
significant bit strings, perform a smaller, local bit-reversed
permutation of size $\frac{b}{2}$. These recursive bit-reversed
permutations will all be applied to much smaller and contiguous blocks
of memory, thereby improving the cache performance. The second step,
wherein the most significant and least significant bit strings are
reversed corresponds to a matrix transposition, with the most
significant bits corresponding to the rows of the matrix and the least
significant bits corresponding to the columns (using {\tt C}-style
row-order array organization). Since {\tt x} and {\tt y} use equal
numbers of bits, this swap corresponds to a transposition of a square
matrix, which can be performed in place. Furthermore, an optimal
cache-oblivious algorithm (meaning that it is close to the optimal
possible runtime for any hierarchical cache organization) for matrix
transposition is known, which performs the transposition in further
and further subdivisions\cite{prokop:cache}. Thus performing a
bit-reversed permutation can be reduced to smaller bit reversed
permutations on contiguous blocks of memory and a square matrix
transposition ({\bf Figure~\ref{figure:recursive}}).

When the number of bits $b$ is odd, then a single even-odd permutation
can be performed first: Reversing {\tt index = x~y~z} where {\tt z}
consists of a single bit yields {\tt rev(index) = z~rev(y)~rev(x)}. An
even-odd permutation produces {\tt z~x~y}. Applying a $b-1$
bit-reversed permutation when {\tt z}=0 and another $b-1$ bit-reversed
permutation {\tt z}=1 will yield {\tt z~rev(y)~rev(x)}. This even-odd
permutation for preprocessing when $b$ is odd will slightly increase
the runtime in that case (this even-odd permutation is also performed
out of place by using a buffer of size $\frac{n}{2}$).

The runtime of the recursive method is defined by the recurrence $r(b)
= 2 \cdot 2^{\frac{b}{2}} \cdot r(\frac{b}{2}) + 2^b$. This recurrence
has closed form $r(b) = 2^{b-3} \cdot b \cdot c + 2^b \cdot (b-1)$,
where $c$ is a constant. Thus $r(b) \in \Theta(2^b \cdot b) = \Theta(n
\log(n))$.

In spite of the flaws in the unrolled closed form when $b \gg 1$, it
is very efficient for small to moderate-sized problems (\emph{e.g.},
$b \leq 14$, which corresponds to $n \leq 16384$); therefore, it is an
ideal base case for the recursive method. The recursive calls can be
made using template recursion, thereby enabling the compiler to
optimize code shared by these recursive calls. This recursive method
generalizes to a semi-recursive method, which not only calls the
unrolled implementation when the number of bits is below a threshold,
it also calls the unrolled implementation when the recursion depth
passes beyond a threshold. This can allow the compiler greater
optimizations, because all of the template-recursive bit-reversed
permutations can be inlined (for larger recursion depths, compilers do
not always inline all of the code, which can be seen by the much
higher compilation times when decorating the recursive bit-reversed
permutation function with {\tt attribute~(\_\_always\_inline\_\_)},
which forces inlining in {\tt g++} and {\tt clang}). Note that a
semi-recursive method that only allows a single recursion will have
runtime $r(b) = 2 \cdot 2^{\frac{b}{2}} \cdot 2^{\frac{b}{2}} + 2^b$,
because the recursive calls $r(\frac{b}{2})$ will be replaced with the
unrolled method runtime, $2^{\frac{b}{2}}$. Therefore, the runtime of
that semi-recursive method becomes $2\cdot 2^b + 2^b \in \Theta(2^b) =
\Theta(n)$.

An advantage of the recursive method over COBRA is that (at least when
the number of bits $b$ is even, $\frac{b}{2}$ is even, \ldots until
the base case size or recursion limit are reached) it can be performed
completely in place and without a buffer. Because the recursive method
reduced bit-reversed permutation to smaller bit-reversed permutations
(each of which are performed independently and in an in-place manner)
and to a square matrix transposition, the method is well suited to
parallelization via SIMD, multiple cores, or by broadcasting over
GPUs; parallelization of COBRA would require duplicate buffers for
each parallelization in order to prevent race conditions.\newline

The asymptotic runtimes for all methods are shown in {\bf
  Table~\ref{table:theoretical-runtimes}}.

\begin{table}[ht!]
  \centering
  \small
  \scalebox{0.88}{
    \begin{tabular}{c|cccccccc}
      & Stockham & Bitwise & Bytewise & Pair bitwise & COBRA & Unrolled & XOR & Recursive \\
      \hline
      %& \multicolumn{3}{c}{$d$ known at compile time} \\
	  {\bf Runtime} & \multirow{2}{*}{$n \log(n)$} & \multirow{2}{*}{$n \log(n)$} & \multirow{2}{*}{$n \log(n)$} & \multirow{2}{*}{$n \log(n)$} & \multirow{2}{*}{$n + \frac{n}{t} \log(\frac{n}{t})$} & \multirow{2}{*}{$n$} & $n$ or & $n$ or \\ 
          (asymptotic) & & & & & & & $n \log(\log(n))$ & $n \log(n)$\\
	  \hline
    \end{tabular}
  }
  \caption{{\bf Theoretical runtimes.} The asymptotic runtimes for
    each algorithm are given. Note that the algorithms with lowest
    theoretical runtime are not necessarily superior in practice: For
    example, the bitwise and bytewise methods are both $\in \Theta(n
    \log(n))$, but the bytewise method has a superior runtime constant
    because it uses a table to reverse words using 8 bits at a
    time. Likewise, the pair bitwise, COBRA, and recursive methods all
    accesses memory in more contiguous, local fashions and therefore
    have superior cache locality. For the COBRA method, $t$ is the
    buffer size used. The runtime of the inductive XOR method will be
    either $\in \Theta(n)$ (when count leading zeros is $\Theta(1)$)
    or $\Theta(n \log(\log(n)))$ (when count leading zeros cannot be
    performed in hardware, and thus requires $\in
    \Theta(\log(\log(n)))$ steps). The recursive method will require
    $\in \Theta(n \log(n))$ steps in general, but can be sped up to
    $\in \Theta(n)$ steps when using a semi-recursive approach, which
    limits the recursion depth.}
  \label{table:theoretical-runtimes}
\end{table}

\section*{Results}
\begin{figure}[ht!]
\centering
  \includegraphics[width=4.5in]{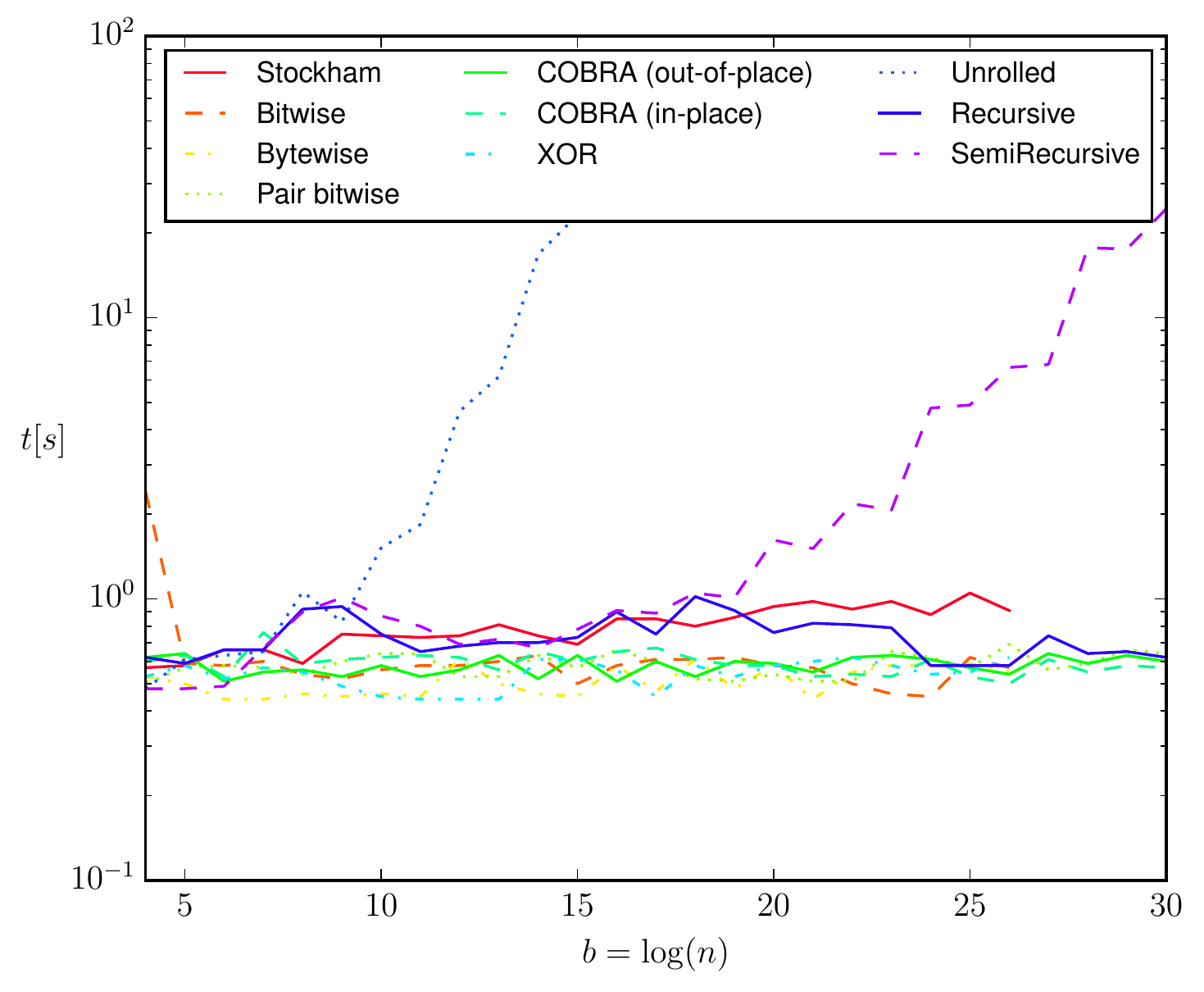}
\caption{{\bf Compile times with {\tt clang++}}. For each method, the
    compile time $t$ (in seconds) for the bit-reversed permutations of various sizes $b$ are
  plotted. Note that the $t$-axis is scaled logarithmically. $n$ is the size of the array 
  that is shuffled, where for each of its $n$ indices $b$ bits have to be reversed. 
  Runtimes for each method are depicted in {\bf Figure~\ref{fig:clang++_runtimes}}.
  Note that UnrolledShuffle has been exluded from compilation for
  $b>16$, since the size of the executables, as well as compile times
  would get unreasonably big.
  \label{fig:clang++_compile_times}
}
\end{figure}

\begin{figure}[ht!]
\centering
  \includegraphics[width=4.5in]{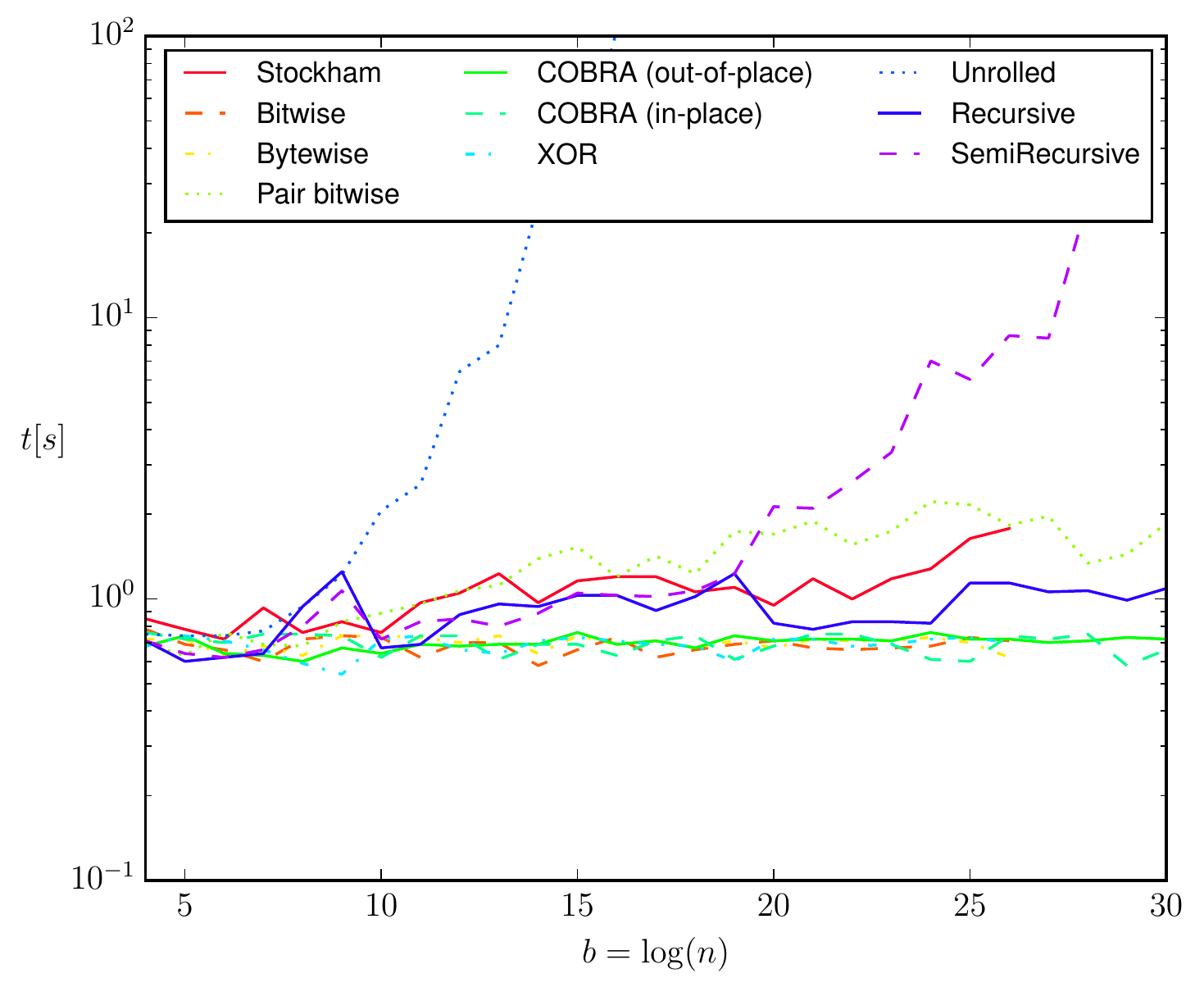}
\caption{{\bf Compile times with {\tt g++}}. For each method, the
    compile time $t$ (in seconds) for the bit-reversed permutations of various sizes $b$ are
  plotted. Note that the $t$-axis is scaled logarithmically. $n$ is the size of the array that 
  is shuffled, where for each of its $n$ indices $b$ bits have to be reversed. 
  Runtimes for each method are depicted in {\bf Figure~\ref{fig:g++_runtimes}}. Note
  that UnrolledShuffle has been exluded from compilation for $b>16$,
  since the size of the executables, as well as compile times would
  get unreasonably big.
  \label{fig:g++_compile_times}
}
\end{figure}

All methods were implemented in {\tt C++11}, making use of template
recursion, along with {\tt constexpr} variables and
functions. Runtimes were compared, benchmarking all methods by
compiling with both {\tt clang++ 3.8.0} and {\tt g++ 6.2.0}. On both
compilers, compilation was optimized with flags {\tt -Ofast
  -march=native -mtune=native}. In order not to risk pointer aliasing
or other interference between different benchmarks, each algorithm and
input size was benchmarked in a separate {\tt main.cpp} file.  Compile
times with {\tt clang++} are plotted in {\bf
  Figure~\ref{fig:clang++_compile_times}} compile times with {\tt g++}
in {\bf Figure~\ref{fig:g++_compile_times}}. All methods were applied
to arrays of type {\tt std::complex<double>}, since our primary
motivation of performing fast bit-reversed permutation lies in
evaluating the methods' suitability for FFT implementation. Note that
individual performances of the methods can vary slightly when other
data types with different sizes (\emph{e.g.}, {\tt int} or {\tt
  float}) are used.

For both the in-place and out-of-place COBRA methods, the buffer size
was optimized for each problem size. The cache-oblivious recursive
method used a base cases of size $b \leq 9$, at which point it
computed the unrolled closed form of bit-reversed permutation of that
size. The runtime was not very sensitive to this choice of base case
size; this is reminiscent of the base case size used in
cache-oblivious matrix transposition, where it is merely used to
amortize out the cost of recursions by ensuring the computational cost
of the base case is non-trivial. 

\begin{figure}[ht!]
\centering
  \includegraphics[width=6in]{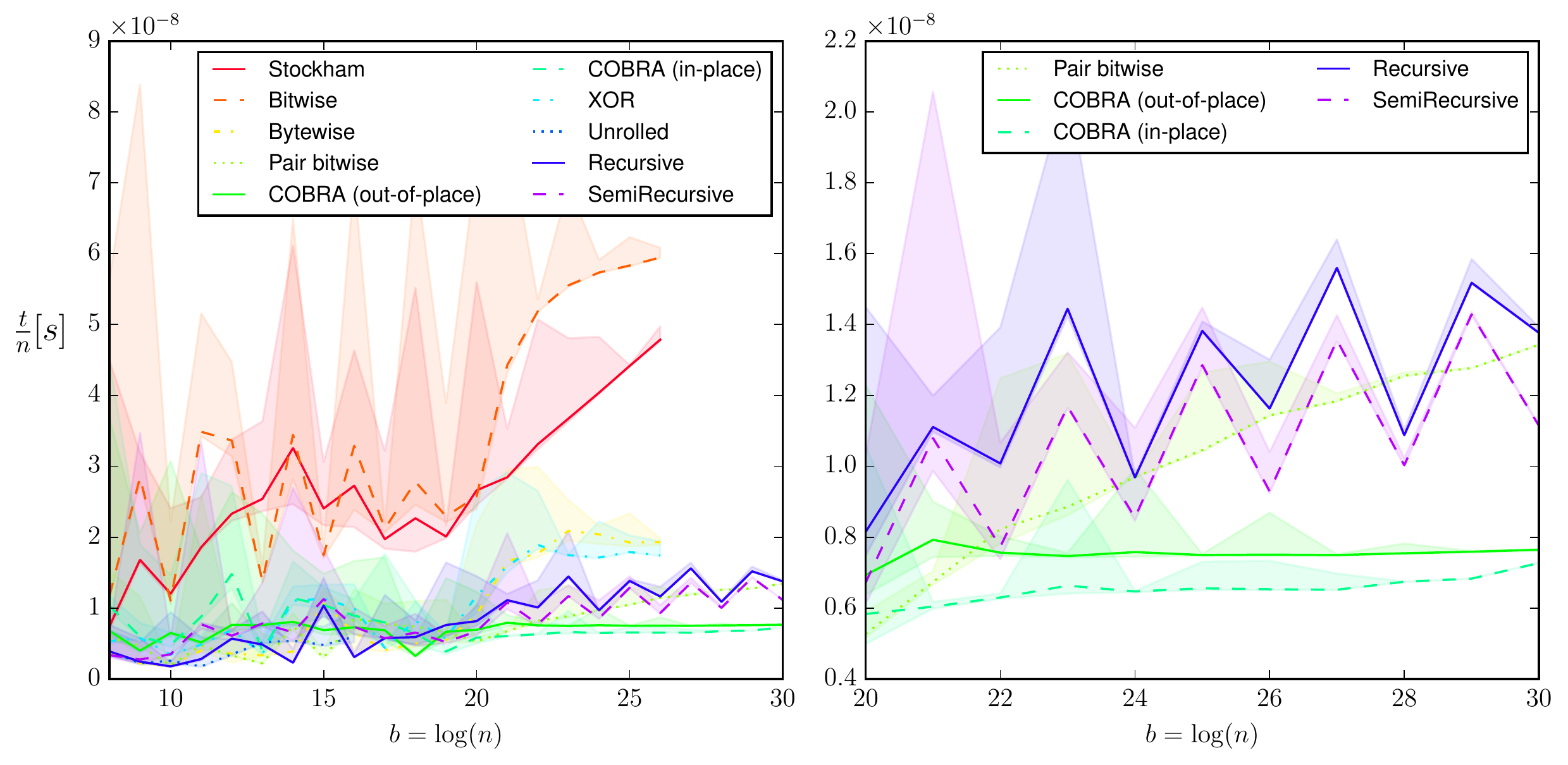}
\caption{{\bf Runtimes per element with {\tt g++}}. For each method,
  bit-reversed permutations of various sizes are performed. For each
  method and on each problem size $b$, 100 replicate runtimes (in seconds)
  were measured, and the average runtime per element (\emph{i.e.},
  elapsed time $t$ divided by $n = 2^b$) are plotted. The shaded areas
  around each series depict the minimum and the maximum runtimes in 
  all of the 100 replicates. The left panel shows all
  methods from $b=8$ to $b=30$ and the right panel shows only the
  highest performance series on larger problems $b=20$ to $b=30$. The
  methods with poor performance on large problem sizes (Stockham,
  bitwise, bytewise and XOR) have been excluded from benchmarking for
  $b>26$.
  \label{fig:clang++_runtimes}	
}
\end{figure}

\begin{figure}[ht!]
\centering
  \includegraphics[width=6in]{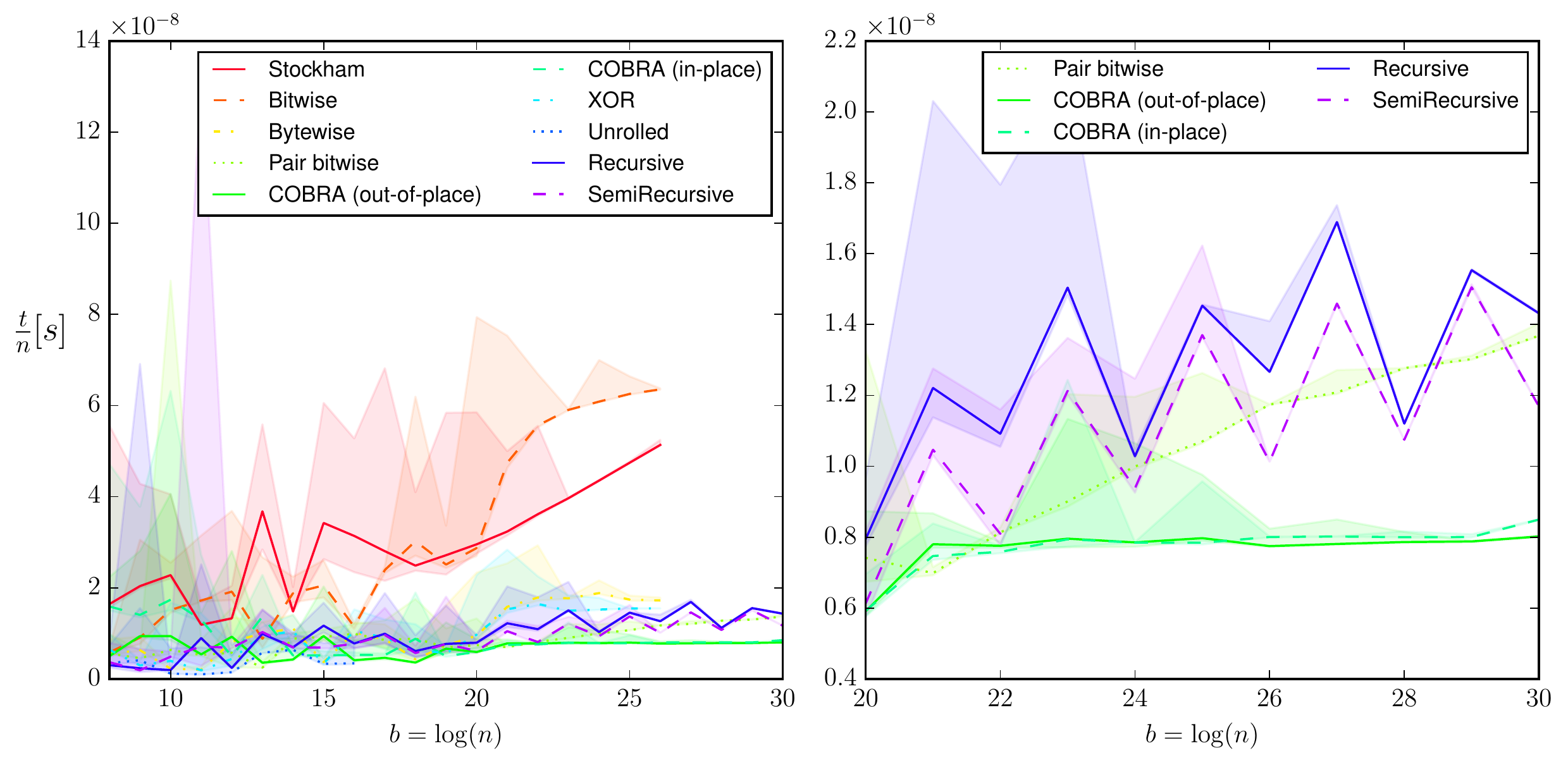}
\caption{{\bf Runtimes per element with {\tt g++}}. For each method,
  bit-reversed permutations of various sizes are performed. For each
  method and on each problem size, 100 replicate runtimes (in seconds)
  were measured, and the average runtime per element (\emph{i.e.},
  elapsed time $t$ divided by $n = 2^b$) are plotted. The shaded areas
  around each series depict the minimum and the maximum runtimes in 
  all of the 100 replicates. The left panel shows all
  methods from $b=8$ to $b=30$ and the right panel shows only the
  highest performance series on larger problems $b=20$ to $b=30$. The
  methods with poor performance on large problem sizes (Stockham,
  bitwise, bytewise and XOR) have been excluded from benchmarking for
  $b>26$.
  \label{fig:g++_runtimes}	
}
\end{figure}

Benchmarked problem sizes ranged from from $n = 2^8$ (which requires
$\approx 4$KB to store) to $n = 2^{30}$ (which requires $\approx
16.4$GB to store). All measurements were averaged over 100 runs, and
runtimes are reported in seconds per element (\emph{i.e.}, the total
elapsed time for the bit-reversed permutation divided by $n$). The CPU
specifications of the computer used for benchmarking are listed in
{\bf Table~\ref{table:cpu_spec}}. The runtimes for all tests, compiled
with {\tt clang++} and {\tt g++} are shown in {\bf
  Figure~\ref{fig:clang++_runtimes}} and {\bf
  Figure~\ref{fig:g++_runtimes}}, respectively.

\begin{figure}[ht!]
\centering
  \includegraphics[width=6in]{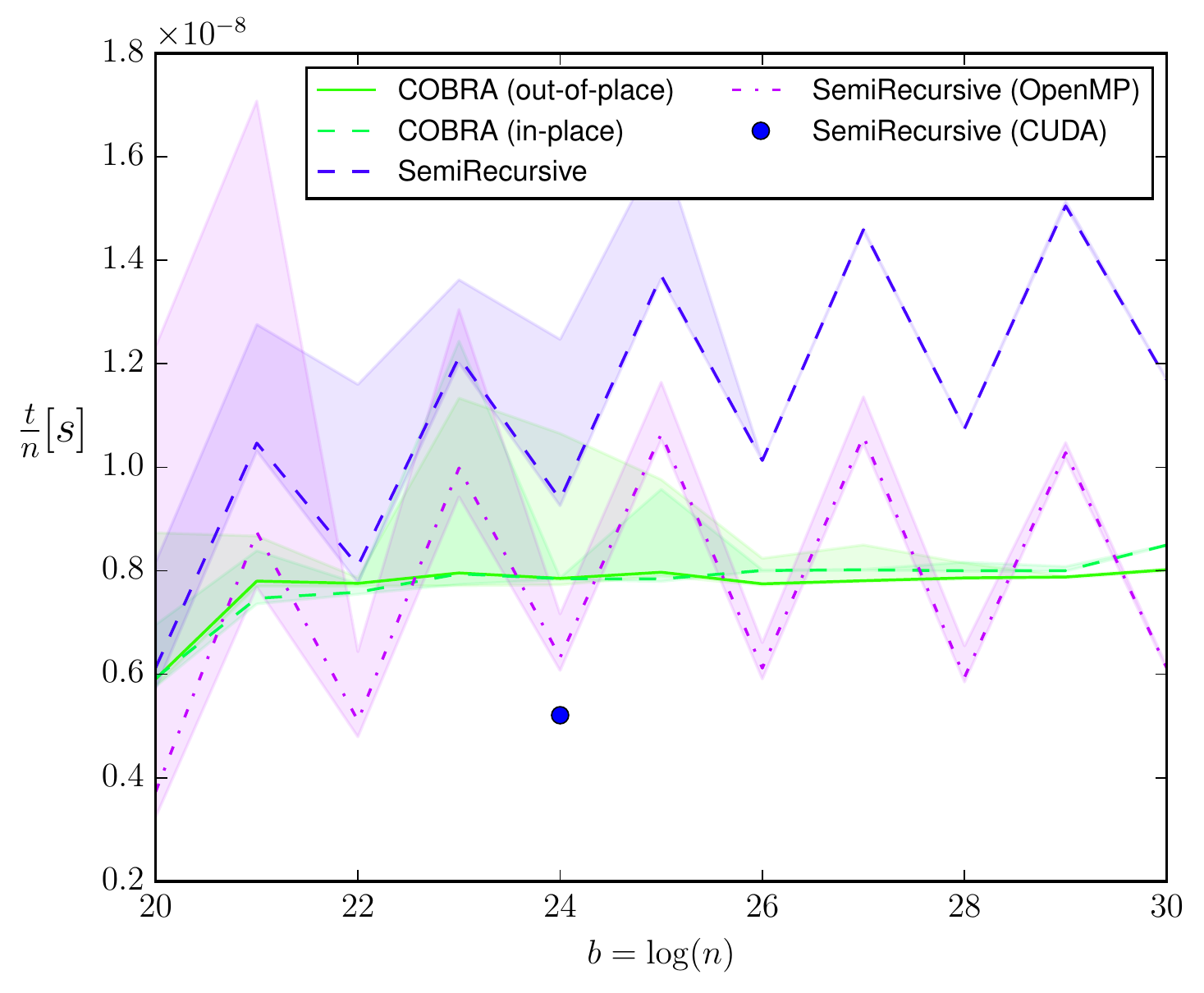}
\caption{{\bf Performance gains from parallelization}. Benchmarks were
  repeated from the right panel of {\bf
   Figure~\ref{fig:g++_runtimes}}, but with the inclusion of a
  parallelized versions of the semi-recursive method. Because of the
  semi-recursive methods inherently parallel nature, OpenMP easily
  results in faster performance. Likewise, broadcasting operations
  over the GPU (via CUDA) achieves even greater parallelism.
  \label{fig:g++_parallel_runtimes}
}
\end{figure}

In order to measure its inherently parallelizable nature, parallel
versions of the semi-recursive method were also implemented and
compared: The first of these multiple cores via OpenMP and the {\tt
  -fopenmp} compiler option. This OpenMP implementation uses {\tt
  \#pragma omp parallel for} to allow the recursive calls (which will
be implemented via the unrolled closed form since only one recursion
is permitted) to be run in parallel. The second parallel version
adapts the semi-recursive method to use the GPU via the CUDA
toolkit. This GPU implementation permits two recursions and is
hard-coded for $b=24$ (it requires the indices swapped to be stored in
an array, which is broadcast over the GPU). $b=24$ was chosen because
it was the largest problem that would wholly fit on the GPU such that
$b$ is divisible by 4 (problems divisible by 4 can be expanded by two
recursions without performing an even-odd permutation as
preprocessing). This CUDA implementation also performs the in-place
transposition on the GPU\cite{harris:cuda}, and therefore parallelizes
both the recursive calls and the transposition (this has an added
benefit of only needing to move data to the GPU once). Both the OpenMP
and the GPU parallel versions were compiled with {\tt g++} (and the
GPU version used {\tt nvcc}), because of limited support to-date for
both OpenMP and CUDA in {\tt clang++}. These parallel versions are
compared to the best-performing single-threaded methods in {\bf
  Figure~\ref{fig:g++_parallel_runtimes}}.

\begin{table}[ht!]
  \centering
  \begin{tabular}{cccccc} L1d &  L1i &    L2 &      L3 & MAX\_SPEED &   RAM \\
\midrule
 32K &  32K &  256K &  15360K &    3.8Ghz &  65GB \\
\end{tabular}
\caption{ {\bf CPU specification used for benchmarking.} The size of
  the L1 data cache, the L1 instruction cache, the L2 cache, the L3
  cache, and the clockspeed as well as the RAM-size of the computer
  used for benchmarking are shown. To relate this to the
  benchmark-results we have shown, note that $32$K can hold an
  array of $n=2^{11}$ elements of type {\tt std::complex<double>},
  $256$K can hold $n=2^{14}$ elements, and $15360$K can hold $n=2^{20}$
  elements. The $65$GB of RAM can hold $n=2^{31}$ elements. 
  \label{table:cpu_spec}
}
\end{table}

\section*{Discussion}
With both {\tt g++} and {\tt clang++}, the Stockham and naive bitwise
shuffling methods have roughly similar performance to one another and
are the least efficient methods investigated here. The Stockham method
is an out-of-place procedure (thus increasing the burden on the cache)
and performs more swap operations than the bitwise shuffle; however,
the Stockham method accesses data by even and odd elements and thus
access the data in a roughly contiguous fashion, while the bitwise
method accesses the data in a less contiguous fashion because it
accesses index {\tt rev(i)} at every iteration.

The bytewise table method and the XOR method achieve similar
performance to one another. Both methods scale considerably better
than the Stockham and bitwise methods. Both the bytewise table method
and the XOR method achieve faster bit reversal, but neither achieves a
contiguous memory access pattern; therefore, neither is very cache
performant. The proposed XOR method may still have use in areas where
memory is at a premium (\emph{e.g.}, embedded systems), since it
achieves similar performance to the byte table without storing or
accessing a table of $256$B.

For small problems, the unrolled approach achieves essentially optimal
performance, seeing as the entire array can be fit into the L1 cache
and there is no overhead for looping, bit reversal, or branch
statements. Furthermore, the operations can be compiled to use
immediate mode addressing in the assembly code, meaning that the
indices swapped are natively encoded into the assembly instructions
themselves and need not be encoded in a separate array. The runtime
benefit of this unrolled method does not extend to larger problems
(because the array doesn't fit into the L1 or L2 caches) and the
compile time of the unrolled method becomes very expensive (roughly
100 seconds when $b=16$ for both {\tt g++} and {\tt clang++}).

The most performant algorithms on large problems are those designed
with the cache in mind: this includes the pair bitwise method, the
COBRA method (both in-place and out-of-place), and the recursive
method and its semi-recursive variant. The recursive and
semi-recursive methods benefit when applied to problems with an even
number of bits, because no even-odd permutation need be performed as a
preprocessing step (this results in a sawtooth pattern in the runtimes
of large problems). The semi-recursive method achieves slightly
greater performance than the recursive method, but at the cost of an
increased compilation time (between 10 to 13 seconds when $b=28$ using
both {\tt g++} and {\tt clang++}). For smaller problems, the
out-of-place COBRA method is less efficient than its in-place variant,
because of the greater burden on the cache from using twice as much
memory. However, for larger problems, a greater performance is
achieved by the out-of-place COBRA method, because the out-of-place
method copies values via a buffer, while the in-place method swaps
values via the buffer; therefore the in-place method must copy to the
buffer, then swap between the buffer and the array, then swap those
resulting changes to the buffer back into the array.

The pair bit reversal method performs more swap operations, but with
greater locality and without the use of a result buffer (as is
required by out-of-place algorithms like Stockham). The pair bitwise
method, the COBRA methods, and the recursive and semi-recursive
methods perform roughly similarly for large problems; the out-of-place
COBRA method performs slightly better for larger problems, but its
performance gains are muted when the matrix buffer size is not
optimized for the CPU. There are small increases in the runtime per
element at cache boundaries, most notably at the L3 cache boundary
encountered at $b=18$ for in-place methods and at $b=17$ for
out-of-place methods.

Unlike the COBRA methods, which achieve their performance via a matrix
buffer, the recursive and semi-recursive methods are inherently cache
efficient and use no such buffer; therefore, they are well suited to
parellelism. {\bf Figure~\ref{fig:g++_parallel_runtimes}} shows the
strong performance benefit that can be achieved by either adding
coarse-grain parallelism via OpenMP or by adding fine-grain
parallelism with the GPU. The recursions do not need to access any
information from other elements, so are perfectly suitable for
concurrency (including the fact that modern architectures sometimes
include separate L1 caches for each hardware core). This inherent
parallelizable quality to the recursive and semi-recursive methods
could likely also be exploited to greater effect by further optimizing
the GPU code.

At $b=24$ in {\bf Figure~\ref{fig:g++_parallel_runtimes}}, the runtime of
the fastest non-parallelized method requires just under $8 \times
{10}^{-9}$ seconds per element, or roughly $0.13$ seconds total. In
comparison, the OpenMP variant of the semi-recursive method requires
roughly $0.1$ seconds, and the GPU variant of the semi-recursive
method requies $0.08$ seconds. As a point of reference, the {\tt
  numpy} FFT (which uses the {\tt FFTPACK} library) takes roughly 2
seconds for problems of that size. The naive bitwise approach requires
roughly 1 second, indicating that around 50\% of the total FFT runtime
would be spent in the bit-reversed permutation, meaning that the
speedup from using these faster bit-reversed permutaion methods is
non-trivial; although the FFT butterfly code performs more
sophisticated manipulations of complex numbers, it does so in a fully
contiguous fashion, making the bit-reversed permutation more important
than it may seem. Furthermore, using this semi-recursive method to
perform the full FFT on the GPU would not only benefit from performing
small FFT butterfly operations in paralel, it would also substantially
reduce the runtime of the bit-reversed permutation by avoiding the
cost of copying data to and from the GPU.

Aside from the ability to parallelize the recursive methods, their
benefit comes from the fact that they achieve high performance without
tuning for a specific cache architecture (which we have found
substantially influences the efficiency of the COBRA
methods). Specifically, the recursive method is cache-oblivious. When
paired with an optimal cache-oblivious method for matrix
transposition\cite{prokop:cache}, it guarantees fairly contiguous
memory accesses without any knowledge of the cache architecture. Like
the recursive method, the semi-recursive strategy also does not use a
buffer with size tuned for cache performance; however, the
semi-recursive is not cache-oblivious, because large problems may be
reduced to smaller bit-reversed permutation problems that still do not
fit in the cache. The recursive method is not only interesting for
future research that would investigate the performance benefit from
parallelizing it, the fact that it achieves high performance without
architecture-specific cache tuning suggests that it is a promising
foothold in the search for an optimal cache-oblivious bit-reversed
permutation algorithm.

\section*{Availability}
All {\tt C++11} source code accompanying this paper, preliminary GPU
implementation of the recursive algorithm, as well as benchmarking
scripts, and the \LaTeX\ code for the paper itself are freely
available (Creative Commons license) at
\url{https://bitbucket.org/orserang/bit-reversal-methods}.\newline

\section*{Acknowledgements}

We are grateful to Thimo Wellner and Guy Ling for their
contributions.\newline 

\noindent This paper grew out of the masters course in Scientific
Computing taught by Oliver Serang. Translations of this paper are also
available in the mother tongues of every student: German, Chinese,
Turkish, and Spanish (these files and their \LaTeX source are linked
by the repository).

\section*{Author Contributions}
The new algorithms and non-parallel implementations in {\tt C++11}
were created by and the project was supervised by O.S. The
Introduction section was written by X.W. The Methods section and
runtime analysis were performed by B.A., X.W., and D.W. The OpenMP
implementation was made by O.S. and the GPU variant was made by
B.A. The benchmarks, figures, and Results sections were created by
C.K. The Discussion section was created by X.W. and C.K. Testing and
optimization were performed by B.A. and D.W. Final benchmarks were
done by C.K., L.I., T.C., and O.S.\newline

% at end of this section:
\noindent Author order was chosen by a secret ballot among the
students.

% bibliography:

\end{document}